\documentclass[twocolumn,superscriptaddress,showpacs,preprintnumbers,amsmath,amssymb]{revtex4}

\usepackage{graphicx,color}
\usepackage{dcolumn}
\usepackage{bm}
\usepackage{CJK}

\newcommand{\lc}{\left<}
\newcommand{\rc}{\right>}
\newcommand{\lr}{\left|}
\newcommand{\rl}{\right|}
\newcommand{\lb}{\left(}
\newcommand{\rb}{\right)}

\begin{document}
\begin{CJK*}{GBK}{song}

\title{Perturbative interpretation of relativistic symmetries in nuclei}

\author{Haozhao Liang}
 \affiliation{State Key Laboratory of Nuclear Physics and Technology, School of Physics,
Peking University, Beijing 100871, China}
 \affiliation{Institut de Physique Nucl\'eaire, IN2P3-CNRS and Universit\'e Paris-Sud,
    F-91406 Orsay Cedex, France}

\author{Pengwei Zhao}
 \affiliation{State Key Laboratory of Nuclear Physics and Technology, School of Physics,
Peking University, Beijing 100871, China}

\author{Ying Zhang}
 \affiliation{State Key Laboratory of Nuclear Physics and Technology, School of Physics,
Peking University, Beijing 100871, China}

\author{Jie Meng}
\affiliation{School of Physics and Nuclear Energy Engineering, Beihang University,
              Beijing 100191, China}
 \affiliation{State Key Laboratory of Nuclear Physics and Technology, School of Physics,
Peking University, Beijing 100871, China}
 \affiliation{Department of Physics, University of Stellenbosch, Stellenbosch, South Africa}

\author{Nguyen Van Giai}
\affiliation{Institut de Physique Nucl\'eaire, IN2P3-CNRS and Universit\'e Paris-Sud,
    F-91406 Orsay Cedex, France}

\date{\today}

\begin{abstract}
Perturbation theory is used systematically to investigate the
symmetries of the Dirac Hamiltonian and their breaking in atomic
nuclei. Using the perturbation corrections to the single-particle
energies and wave functions, the link between the single-particle
states in realistic nuclei and their counterparts in the symmetry
limits is discussed. It is shown that the limit of $S-V={\rm const}$
and relativistic harmonic oscillator (RHO) potentials can be
connected to the actual Dirac Hamiltonian by the perturbation
method, while the limit of $S+V={\rm const}$ cannot, where $S$ and
$V$ are the scalar and vector potentials, respectively. This
indicates that the realistic system can be treated as a perturbation
of spin-symmetric Hamiltonians, and the energy splitting of the
pseudospin doublets can be regarded as a result of small
perturbation around the Hamiltonian with RHO potentials, where the
pseudospin doublets are quasidegenerate.
\end{abstract}

\maketitle


It is well known that the spin symmetry (SS) breaking, i.e., the
remarkable spin-orbit splitting for the spin doublets ($n,l,j =
l\pm1/2$), is one of the most important concepts for understanding
the traditional magic numbers (2, 8, 20, 28, ...) in atomic nuclei
\cite{Haxel1949,Mayer1949}. Meanwhile, a new symmetry, the so-called
pseudospin symmetry (PSS) \cite{Arima1969,Hecht1969}, is introduced
to explain the near degeneracy between two single-particle states
with the quantum numbers ($n-1, l + 2, j = l + 3/2$) and ($n, l, j=l
+ 1/2$) by defining the pseudospin doublets
($\tilde{n}=n-1,\tilde{l}=l+1,j=\tilde{l}\pm1/2$). The splittings of
both spin and pseudospin doublets play critical roles in the shell
structure evolutions. Thus, it is a fundamental task to explore the
origin of SS and PSS, as well as the mechanism of their breaking.

Since the suggestion of PSS in atomic nuclei, there have been
comprehensive efforts to understand its origin. Apart from the
rather formal relabeling of quantum numbers, various proposals for
an explicit transformation from the normal scheme to the pseudospin
scheme have been discussed \cite{Bohr1982,Castans1992,Blokhin1995}.
Based on the single-particle Hamiltonian of the oscillator shell
model, the origin of PSS is connected with the special ratio in the
strength of the spin-orbit and orbit-orbit interactions
\cite{Bohr1982}. The relation between the PSS and the relativistic
mean field (RMF) theory \cite{Serot1986} was first noted in
Ref.~\cite{Bahri1992}, in which Bahri \textit{et al.} found that the
RMF theory approximately explains such a special ratio in the
strength of the spin-orbit and orbit-orbit interactions.

As substantial progress, the PSS was shown to be a symmetry of the
Dirac Hamiltonian, where the pseudo-orbital angular momentum
$\tilde{l}$ is nothing but the orbital angular momentum of the lower
component of the Dirac spinor, and the equality in magnitude but
difference in sign of the scalar potential $S(\mathbf{r})$ and
vector potential $V(\mathbf{r})$ was suggested as the exact PSS
limit by reducing the Dirac equation to a Schr\"{o}dinger-like
equation \cite{Ginocchio1997}. As a more general condition,
$d(S+V)/dr=0$ can be approximately satisfied in exotic nuclei with
highly diffuse potentials \cite{Meng1998,Meng1999}. Meanwhile, based
on this limit, the pseudospin SU(2) algebra was established
\cite{Ginocchio1998}, and the specific node structures of the
pseudospin doublets were illuminated \cite{Leviatan2001}.
Furthermore, the Dirac Hamiltonian with spin and pseudospin SU(2)
symmetries can also be derived in supersymmetric (SUSY) patterns
\cite{Leviatan2004}. However, since there exist no bound nuclei
within $S+V={\rm const}$, the non-perturbative nature of PSS in
realistic nuclei has been presented in
Refs.~\cite{Alberto2002,Lisboa2010,Ginocchio2011}, which is also
related to the consideration of the PSS as being a dynamical
symmetry \cite{Alberto2001}.

On the other hand, the relativistic harmonic oscillator (RHO)
potentials were used to understand the origin of PSS
\cite{Chen2003,Lisboa2004}. Subsequently, the spin and pseudospin
U(3) algebra was established in the Dirac Hamiltonian with RHO
potentials \cite{Ginocchio2005PRL,Ginocchio2011}. Recently, Typel
pointed out that the Hamiltonian with spin U(3) symmetry is one of
the simplest cases where the pseudospin symmetry-breaking potential
derived in the SUSY framework vanishes \cite{Typel2008}. Meanwhile,
Marcos \textit{et al.} commented that the quasi-degeneracy of the
pseudospin doublets in realistic nuclei can be considered as the
breaking of their degeneracy in the Dirac Hamiltonian with RHO
potentials \cite{Marcos2008}.

In this Rapid Communication, the perturbation theory will be used
for the first time to investigate the symmetries of the Dirac
Hamiltonian and their breaking in realistic nuclei. The perturbation
corrections to the single-particle energies and wave functions will
be accurately calculated numerically. In this way, the link between
the single-particle states in realistic nuclei and their
counterparts in the symmetry limits will be constructed explicitly.

\begin{figure}
\includegraphics[width=0.30\textwidth]{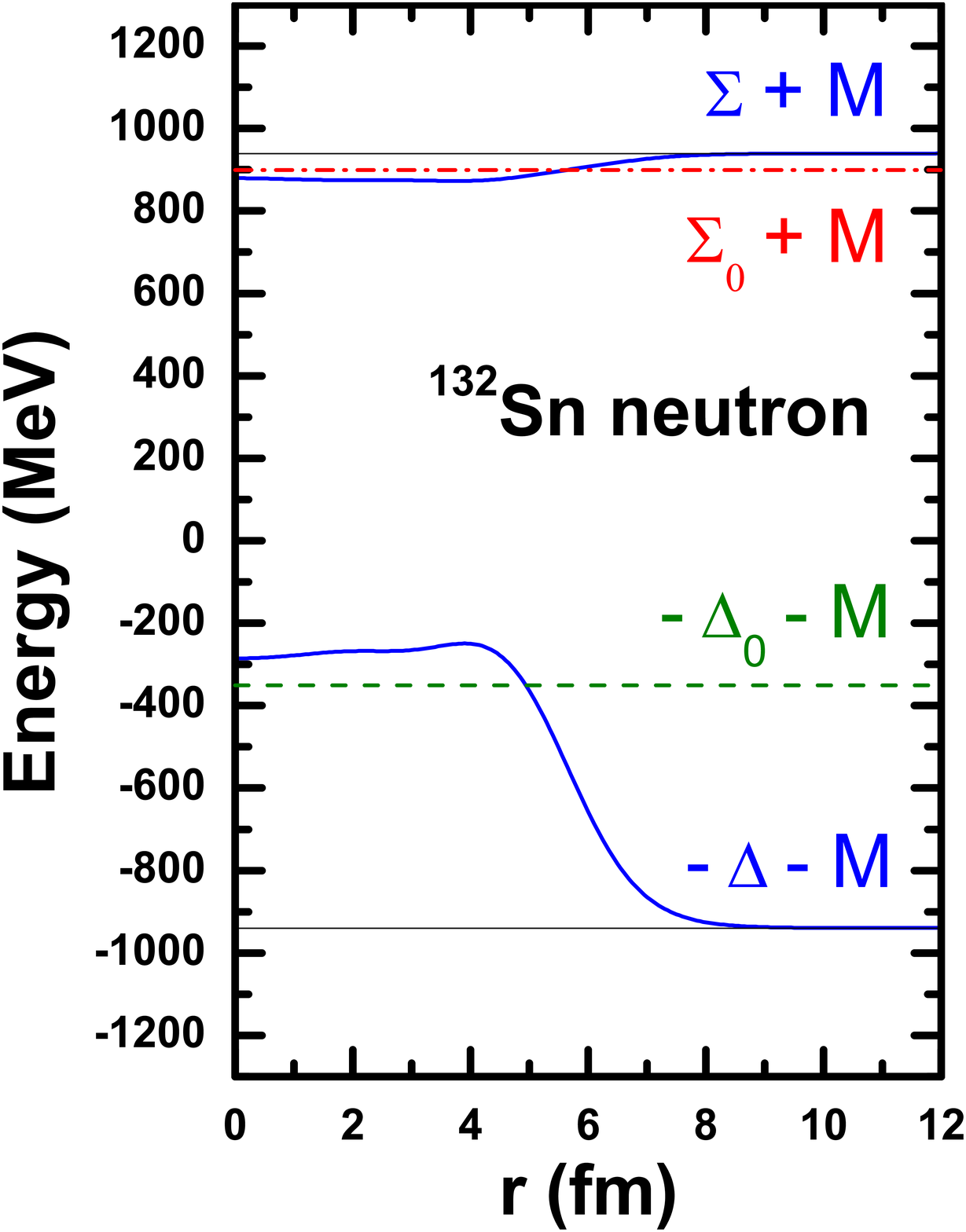}
\caption{(Color online) Single-particle potentials for neutrons in
the nucleus $^{132}$Sn. The self-consistent potentials calculated by
the RMF theory with PK1 \cite{Long2004} are shown as solid lines.
The potentials $-\Delta_0-M$ in $H_0^{\rm SS}$ and $\Sigma_0+M$ in
$H_0^{\rm PSS}$ are illustrated as dashed, and dash-dotted lines,
respectively.
    \label{Fig1}}
\end{figure}

Assuming the spherical symmetry, the radial Dirac equations can be
cast in the form of
\begin{equation}\label{EQ:Dirac}
    H\Psi =E\Psi
\end{equation}
with
\begin{equation}\label{EQ:H}
    H =
    \left(\begin{array}{cc}
        \Sigma(r)+M & -\frac{d}{dr}+\frac{\kappa}{r} \\
        \frac{d}{dr}+\frac{\kappa}{r} & -\Delta(r)-M
    \end{array}\right),~
    \mbox{and}~
    \Psi =
    \left(\begin{array}{c}
        G(r) \\ F(r)
    \end{array}\right),
\end{equation}
where $\Sigma(r)=S(r)+V(r)$ and $\Delta(r)=S(r)-V(r)$ denote the
combinations of the scalar and vector potentials, and $\kappa$ is
defined as $\kappa=(l-j)(2j+1)$. Taking the nucleus $^{132}$Sn as an
example, the potentials $\Sigma(r)$ and $\Delta(r)$ for neutrons
calculated by the self-consistent RMF theory with the effective
interaction PK1 \cite{Long2004} are shown in Fig.~\ref{Fig1}. It is
generally found that $\Sigma(r) \sim 70$~MeV and $\Delta(r) \sim
700$~MeV in the realistic nuclei.

Following the idea of Rayleigh-Schr\"odinger perturbation theory,
the Dirac Hamiltonian $H$ in Eq.~(\ref{EQ:H}) can be split as
\begin{equation}
    H = H_0 + W,
\end{equation}
or equivalently
\begin{equation}
    H_0 = H - W,
\end{equation}
where $H_0$ leads to the exact spin (pseudospin) symmetry and $W$ is
identified as the corresponding symmetry-breaking potential. The
condition
\begin{equation}\label{EQ:criterion}
    \left|\frac{W_{mk}}{E_k-E_m}\right|\ll 1
    \quad\mbox{for}\quad m\neq k,
\end{equation}
where $W_{mk}=\left< \Psi_m \right> W \left< \Psi_k \right>$,
determines whether $W$ can be treated as a small perturbation and
governs the convergence of the perturbation series
\cite{Greiner1994}.

In the case of the spin and pseudospin SU(2) symmetry limits shown
in Ref.~\cite{Leviatan2004}, the Dirac Hamiltonian with exact
symmetries reads
\begin{subequations}\label{EQ:HSU2}
\begin{align}
    H_0^{\rm SS}&=
    \lb\begin{array}{cc}
        \Sigma+M & -\frac{d}{dr}+\frac{\kappa}{r} \\
        \frac{d}{dr}+\frac{\kappa}{r} & -\Delta_0-M
    \end{array}\rb,\\
    H_0^{\rm PSS}&=
    \lb\begin{array}{cc}
        \Sigma_0+M & -\frac{d}{dr}+\frac{\kappa}{r} \\
        \frac{d}{dr}+\frac{\kappa}{r} & -\Delta-M
    \end{array}\rb,
\end{align}
\end{subequations}
whose eigenenergies are denoted as $E_0$ in general, and the
corresponding symmetry-breaking potentials are
\begin{equation}
W^{\rm SS}=
    \lb\begin{array}{cc}
        0 & 0 \\
        0 & \Delta_0-\Delta
    \end{array}\rb,\qquad
    W^{\rm PSS}=
    \lb\begin{array}{cc}
        \Sigma-\Sigma_0 & 0 \\
        0 & 0
    \end{array}\rb.
\end{equation}

In contrast to adopting the Schr\"{o}dinger-like equations in the
previous studies
\cite{Ginocchio1997,Marcos2001,Alberto2001,Alberto2002}, it is
clearly shown that the operators $H$, $H_0$ and $W$ used in the
present calculations are all Hermitian, and they do not contain any
singularity. This allows us to perform the order-by-order
perturbation calculations. In addition, it should also be noticed
that only $W$ corresponds to the symmetry-breaking potential within
the present decomposition, thus the ambiguity caused by the strong
cancellations among the different terms in the Schr\"{o}dinger-like
equations can also be avoided. Therefore, the present method can
provide a clear and quantitative way for investigating the
perturbative nature of SS and PSS. This method can be universally
applied to the cases that the nature of the symmetry is either
perturbative or non-perturbative. For the case where the nature of
the symmetry is perturbative, the link between the single-particle
states in realistic nuclei and their counterparts in the symmetry
limits can be constructed. For the case where the nature of the
symmetry is non-perturbative, the divergence of the perturbation
series can be found explicitly.

In the present calculations, as illustrated with dashed and
dash-dotted lines in Fig.~\ref{Fig1}, the constant potentials in
Eqs.~(\ref{EQ:HSU2}) are chosen as $-\Delta_0-M=-350$~MeV and
$\Sigma_0+M=900$~MeV. We have checked that the convergence of the
perturbation calculations is not sensitive to these values.

\begin{figure}
\includegraphics[width=0.40\textwidth]{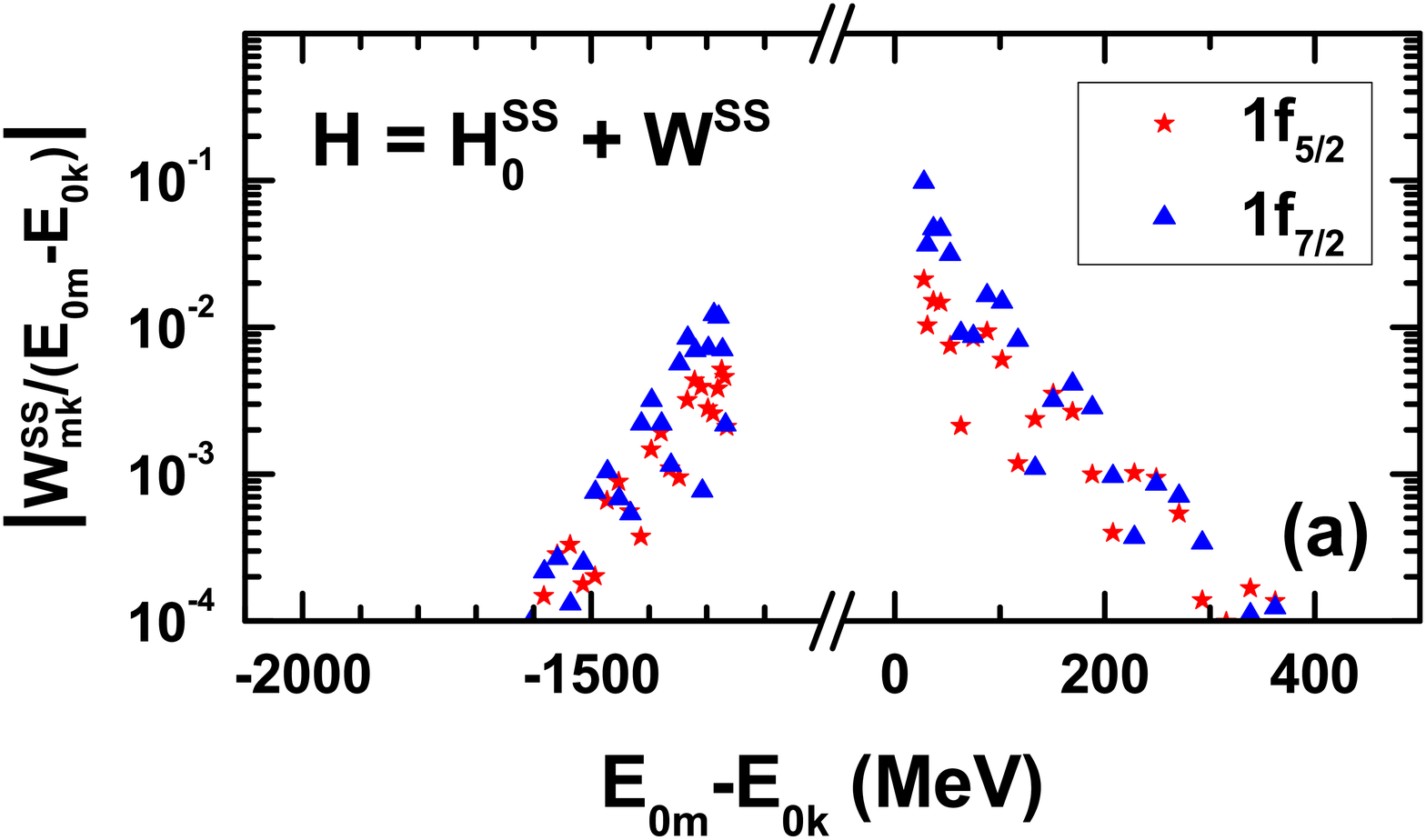}\\
\includegraphics[width=0.40\textwidth]{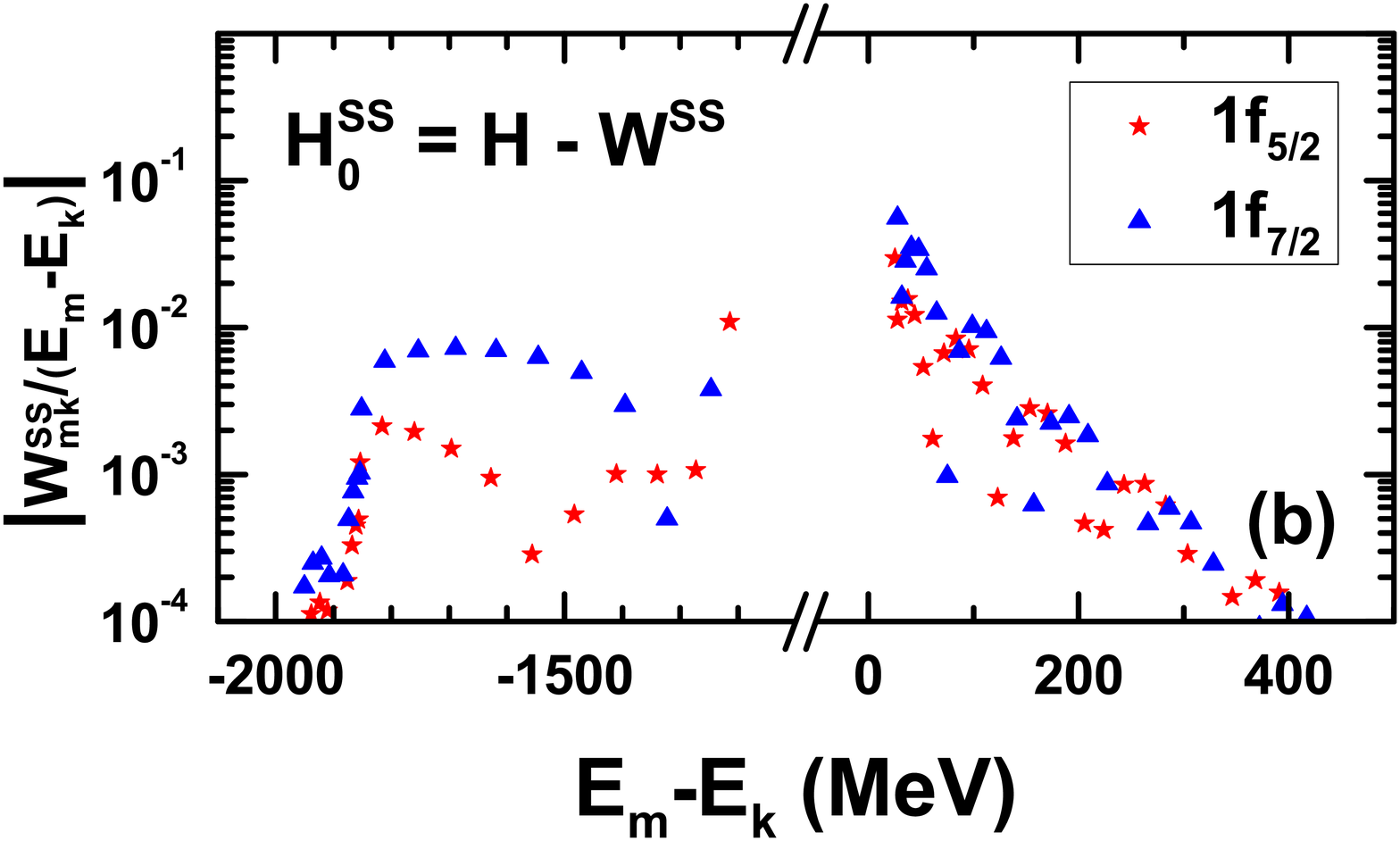}\\
\includegraphics[width=0.40\textwidth]{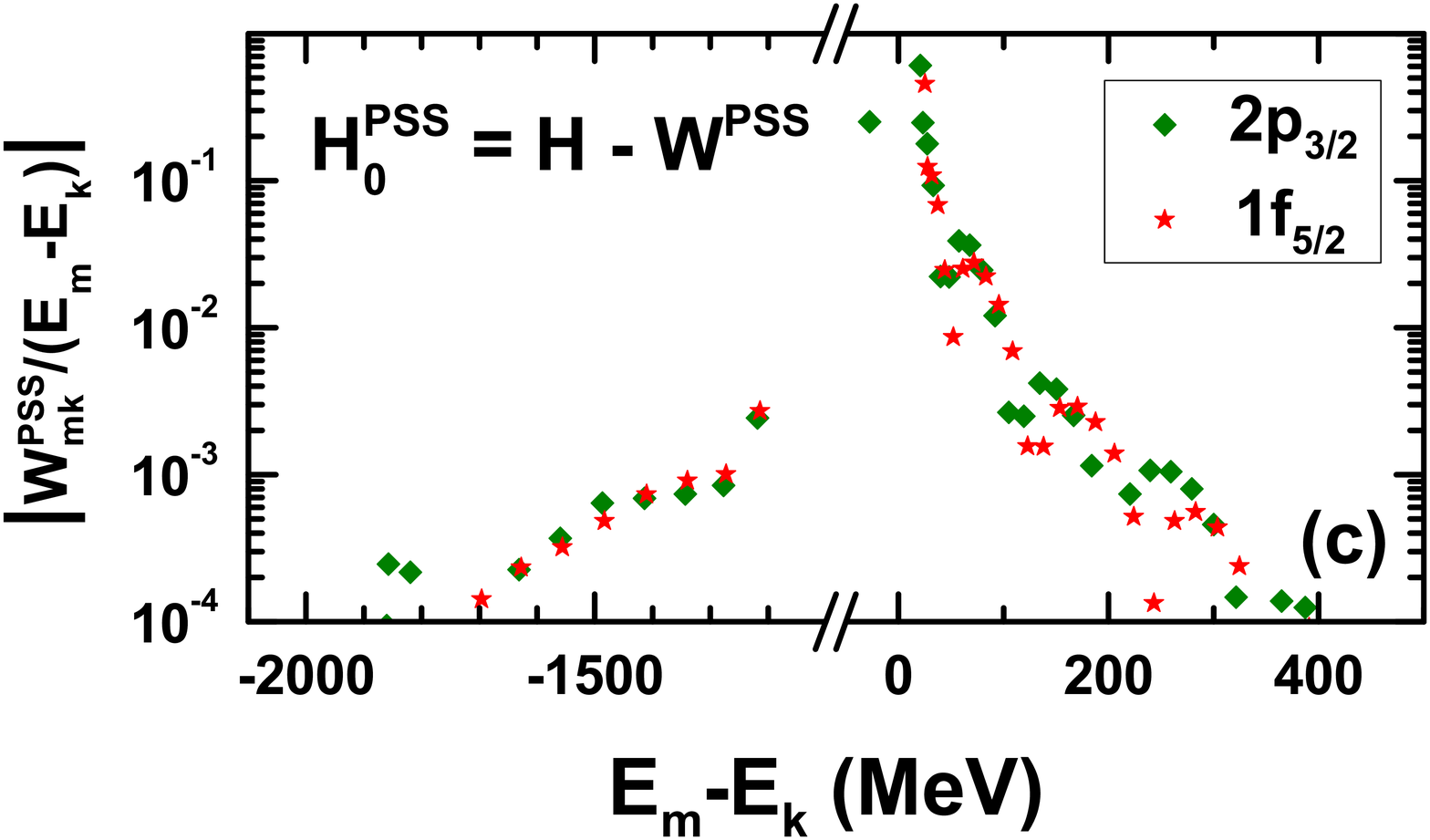}
\caption{(Color online) Values of $\left|W_{mk}/(E_m-E_k)\right|$ vs
the energy differences $E_m-E_k$ for the spin doublets $k=1f$
(panels (a) and (b)) and the pseudospin doublets $k=1\tilde{d}$
(panel (c)). The unperturbed eigenstates are chosen as those of
$H_0^{\rm SS}$ (panel (a)) and $H$ (panels (b) and (c)),
respectively. The single-particle states $m$ include the states in
the Dirac sea and Fermi sea.
    \label{Fig2}}
\end{figure}

In Fig.~\ref{Fig2}, taking the spin doublets $k=1f$ and the
pseudospin doublets $k=1\tilde{d}$ as examples, the values of
$\left|W_{mk}/(E_m-E_k)\right|$ are plotted as functions of the
energy differences $E_m-E_k$. For the SS case, the unperturbed
eigenstates are chosen as those of $H_0^{\rm SS}$ in panel (a),
while the unperturbed eigenstates are chosen as those of $H$ in
panel (b), namely, the former perturbation calculations are
performed from a spin-symmetric Hamiltonian $H_0^{\rm SS}$ to the
realistic Hamiltonian $H$, whereas the latter ones are performed
from $H$ to $H_0^{\rm SS}$. For the PSS case, since there are no
bound states in the pseudospin-symmetric Hamiltonian $H_0^{\rm
PSS}$, the perturbation calculations are performed only from $H$ to
$H_0^{\rm PSS}$ as shown in panel (c). For the completeness of the
single-particle basis, the single-particle states $m$ must include
not only the states in the Fermi sea, but also those in the Dirac
sea. Since the spherical symmetry is adopted, only the states $m$
and $k$ with the same quantum numbers $l$ and $j$ lead to
non-vanishing matrix elements $W_{mk}$. It is seen that the values
of $\left|W_{mk}/(E_m-E_k)\right|$ decrease as a general tendency
when the energy differences $|E_m-E_k|$ increase. From the
mathematical point of view, this property provides natural cut-offs
of the single-particle states in the perturbation calculations.

Although the potentials satisfy $\left|\Delta_0-\Delta\right|\gg
\left|\Sigma-\Sigma_0\right|$, the largest value of
$\left|W_{mk}/(E_m-E_k)\right|$ is roughly 0.10 ($H_0^{\rm SS}$ to
$H$) or 0.06 ($H$ to $H_0^{\rm SS}$) for the SS case, whereas it is
about 0.6 for the PSS case because different components of the Dirac
spinors are involved:
\begin{subequations}
\begin{align}
    W^{\rm SS}_{mk} &= \lc F_m\rl (\Delta_0-\Delta)\lr F_k\rc,\\
    W^{\rm PSS}_{mk} &= \lc G_m\rl (\Sigma-\Sigma_0) \lr G_k\rc,
\end{align}
\end{subequations}
where for the Fermi states the upper component $G(r)\sim O(1)$, and
the lower component $F(r)\sim O(1/10)$. This indicates that the
criterion in Eq.~(\ref{EQ:criterion}) can be well fulfilled for the
SS case, but questionable for the PSS case.

\begin{figure}
\includegraphics[width=0.40\textwidth]{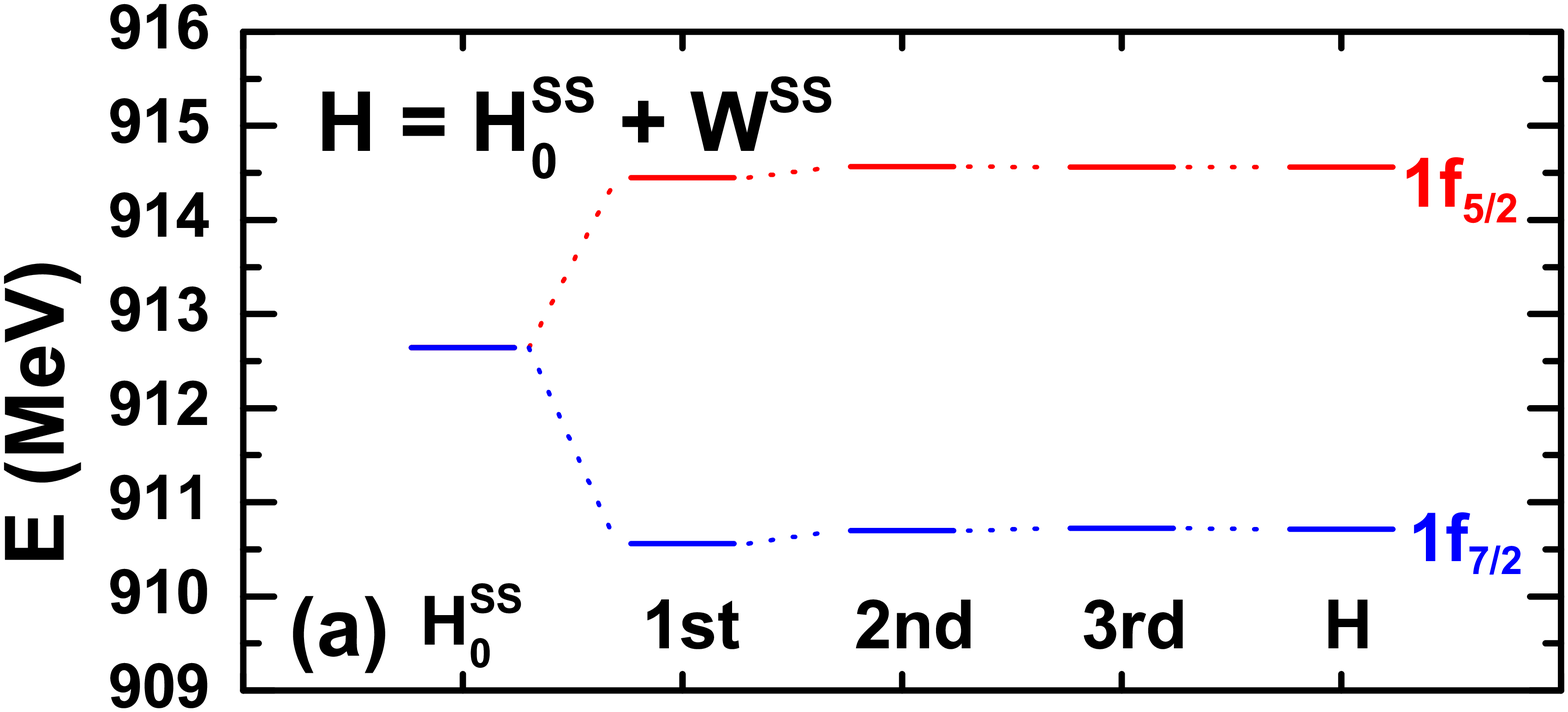}\\
\includegraphics[width=0.40\textwidth]{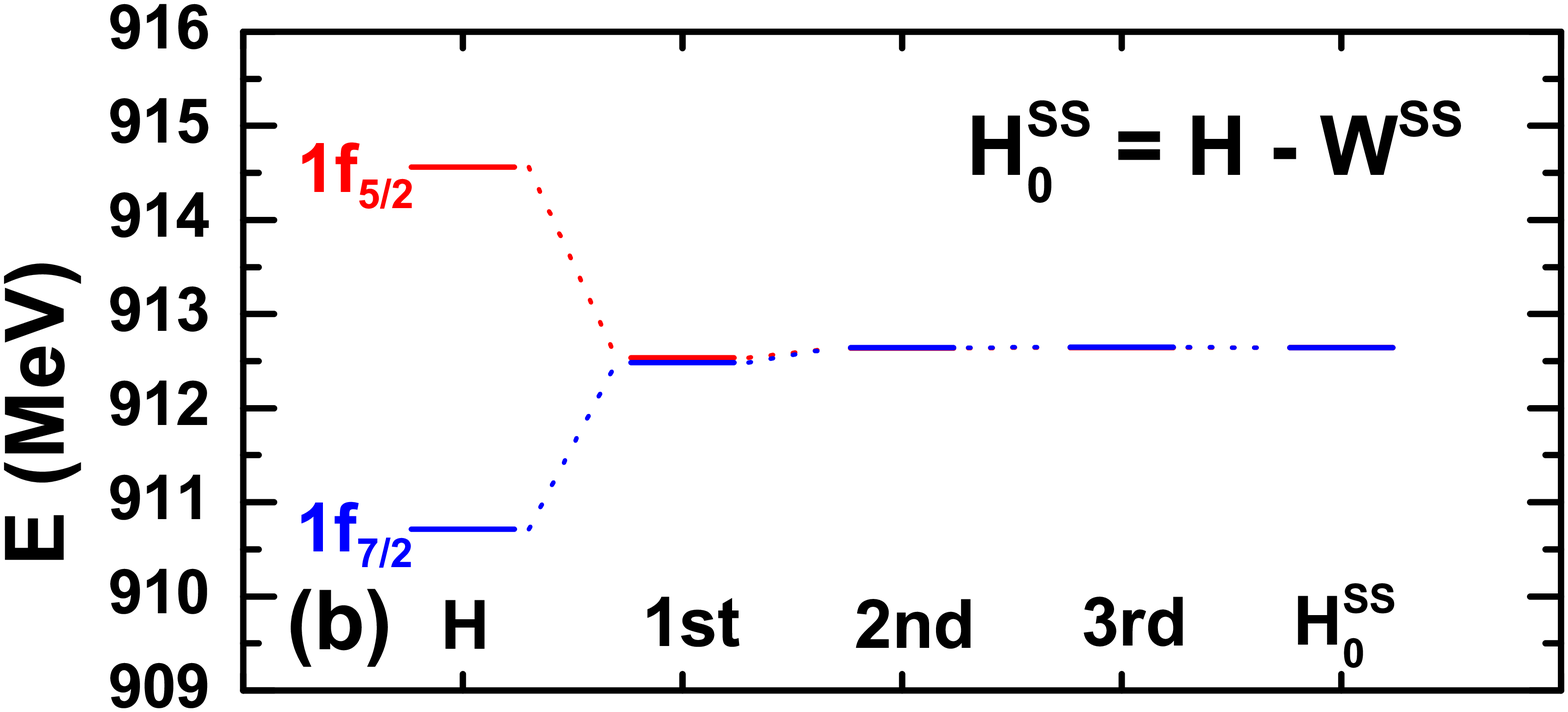}\\
\includegraphics[width=0.40\textwidth]{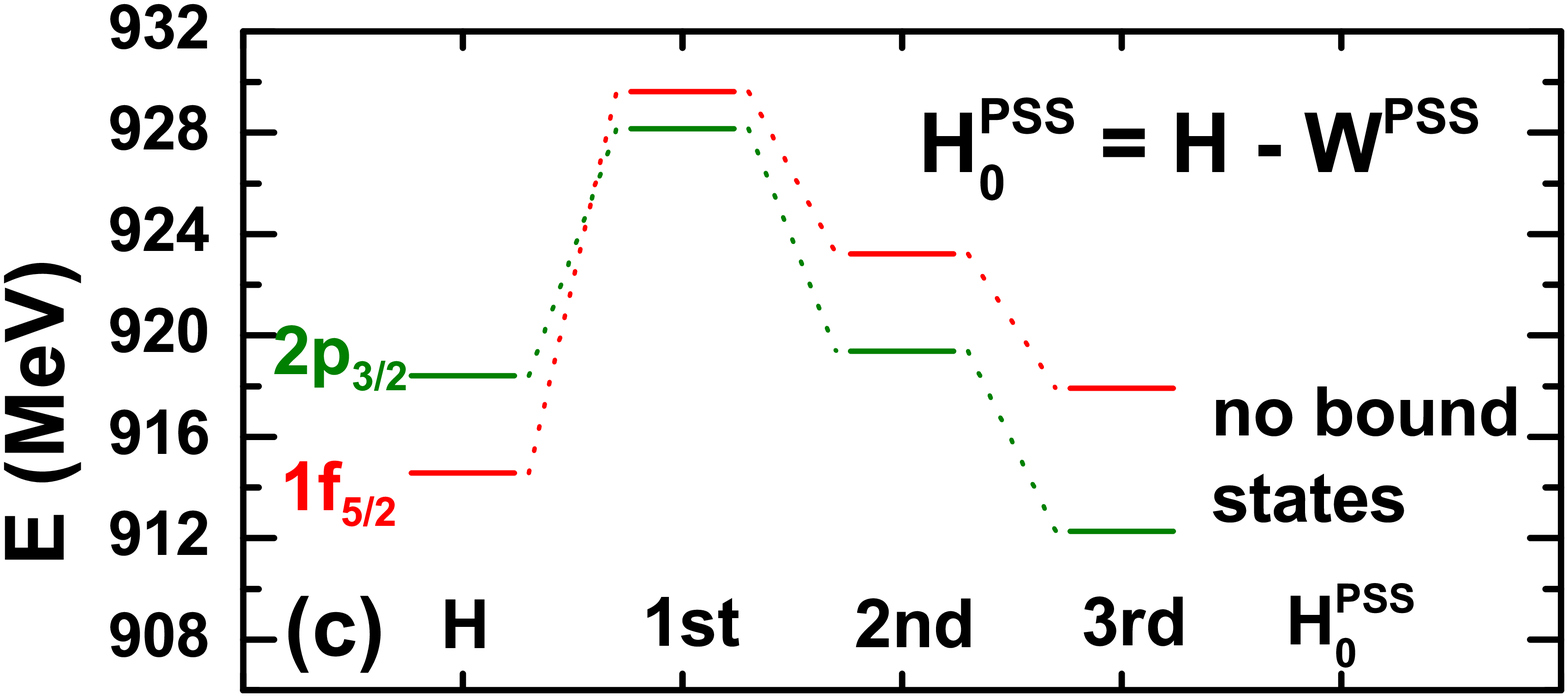}
\caption{(Color online) Single-particle energies of spin doublets
$1f$ (panels (a) and (b)) and pseudospin doublets $1\tilde{d}$
(panel (c)) obtained at the exact symmetry limits, and by the
first-, second-, and third-order perturbation calculations, as well
as those by the RMF theory. The unperturbed eigenstates are chosen
as those of $H_0^{\rm SS}$ (panel (a)) and $H$ (panels (b) and (c)),
respectively.
    \label{Fig3}}
\end{figure}

\begin{figure*}
\includegraphics[width=0.40\textwidth]{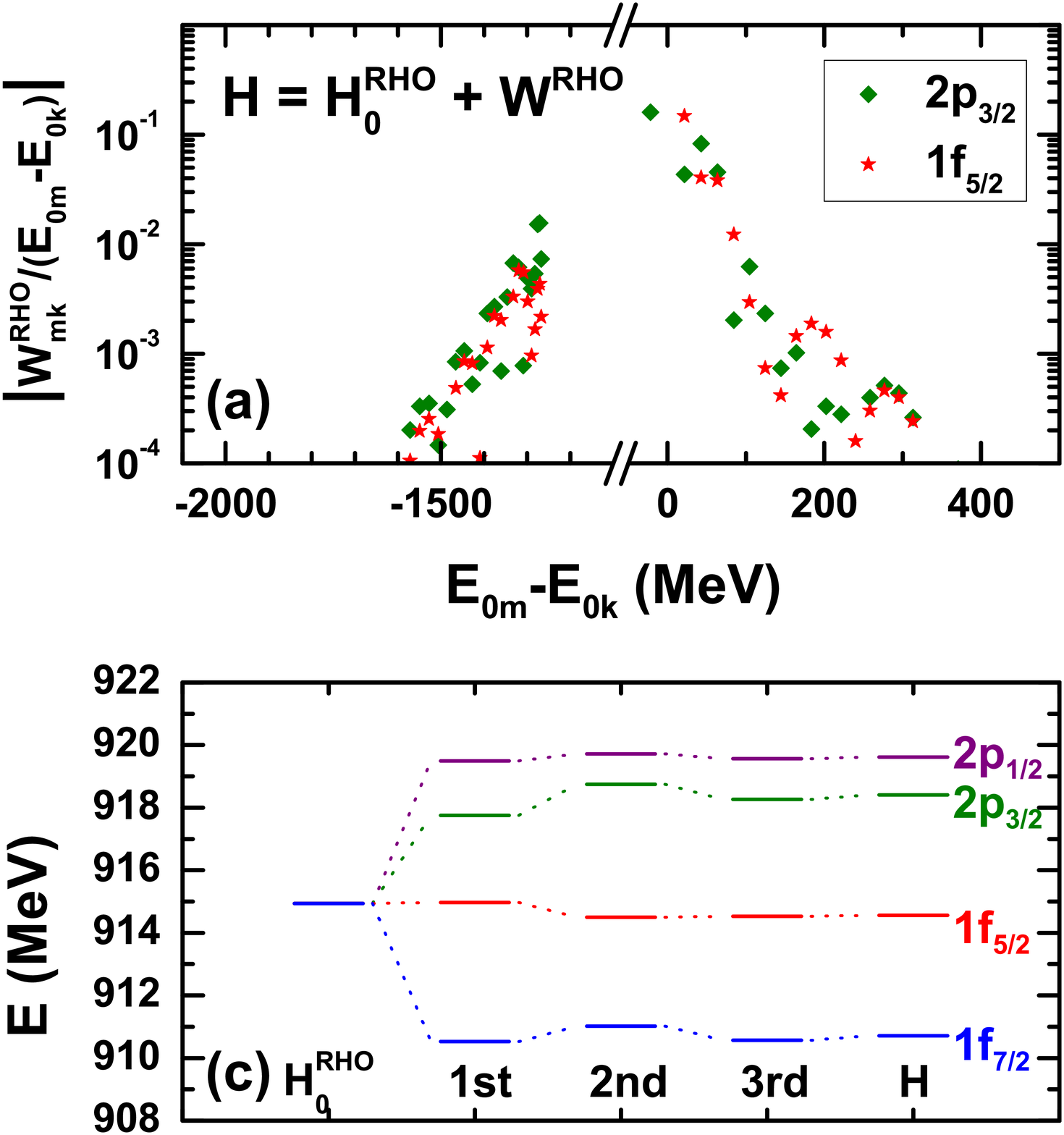}~~~~
\includegraphics[width=0.40\textwidth]{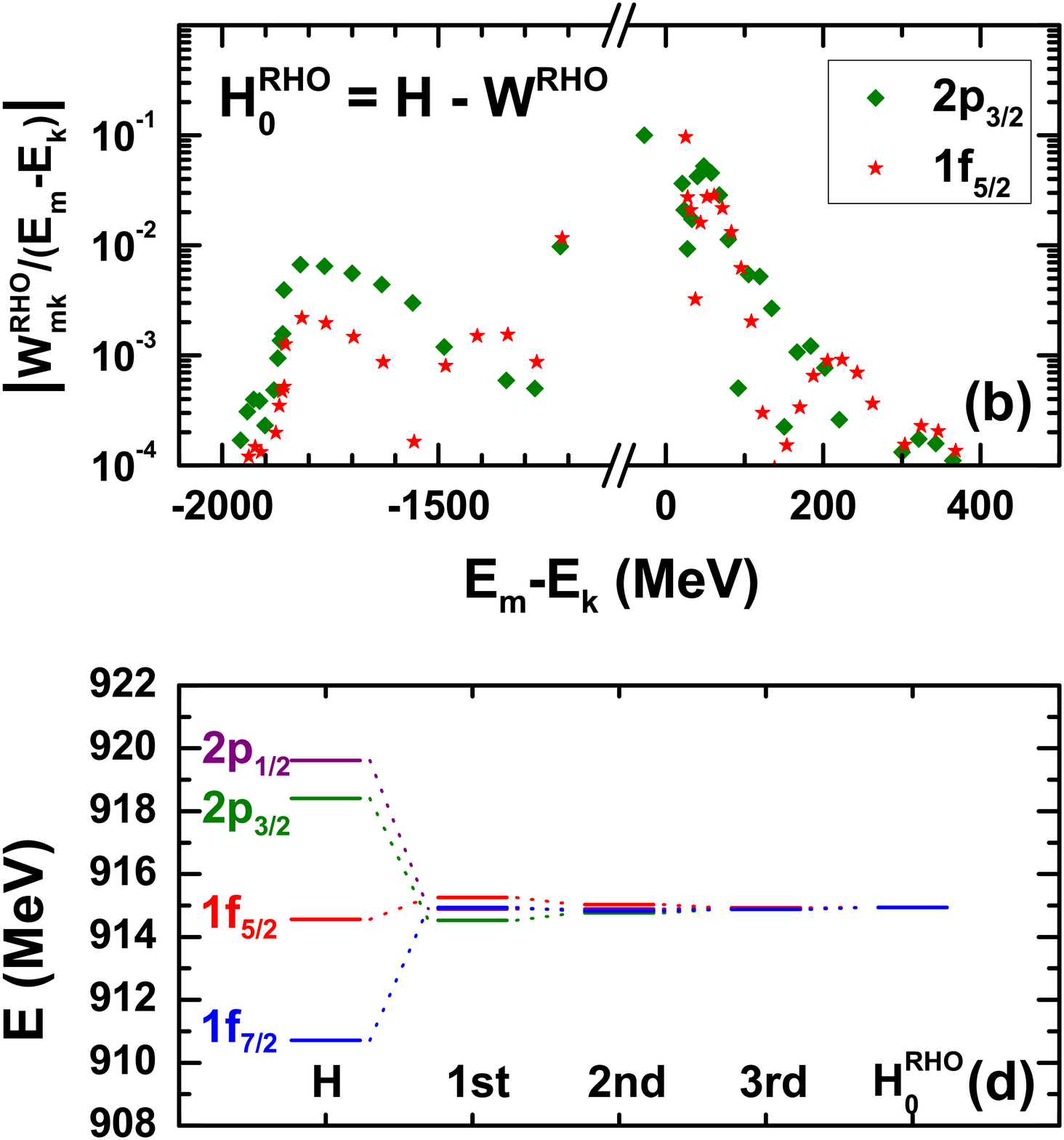}
\caption{(Color online) Upper panels: Same as Fig.~\ref{Fig2}, but
for the case of the RHO potentials. Lower panels: Same as
Fig.~\ref{Fig3}, but for all single-particle states in the $pf$
major shell. The unperturbed eigenstates are chosen as those of
$H_0^{\rm RHO}$ (panels (a) and (c)) and $H$ (panels (b) and (d)),
respectively.
    \label{Fig4}}
\end{figure*}

Let us then examine the perturbation corrections to the
single-particle energies of the spin doublets $1f$ and pseudospin
doublets $1\tilde{d}$. In panel (a) of Fig.~\ref{Fig3}, by choosing
the unperturbed eigenstates as those of $H_0^{\rm SS}$, the
single-particle energies obtained at the exact spin symmetry limit,
and their counterparts obtained by the first-, second-, and
third-order perturbation calculations, as well as those obtained by
the self-consistent RMF theory, are shown from left to right.
Meanwhile, the corresponding results obtained by choosing the
unperturbed eigenstates as those of $H$ are shown in panels (b) and
(c) of Fig.~\ref{Fig3}. It can be seen clearly that the spin-orbit
splitting is well reproduced by the second-order perturbation
calculations as shown in panel (a), and in the reversed way the
energy degeneracy of the spin doublets can be well restored as shown
in panel (b). This verifies that for studying the relationship
between the eigenstates of $H_0$ and $H$ by perturbation theory, it
is equivalent to use the definitions $H=H_0+W$ and $H_0=H-W$. In
contrast, as shown in panel (c), the energy degeneracy of the
pseudospin doublets cannot be restored up to the third-order
perturbation calculations, and there exist no bound eigenstates of
$H_0^{\rm PSS}$. Thus, the link between the pseudospin doublets in
realistic nuclei and their counterparts in the $S+V={\rm const}$
limit is unclear. In addition, choosing the eigenstates of $H$ as
the unperturbed eigenstates, the perturbation corrections to the
single-particle wave functions are also evaluated. It is found that
the identity of the wave functions for the spin doublets,
$G_0(1f_{5/2})=G_0(1f_{7/2})$, can be well reproduced by the
second-order perturbation corrections, but the identity of the wave
functions for the pseudospin doublets $F_0(1f_{5/2})=F_0(2p_{3/2})$,
cannot be fulfilled.

Therefore, from the perturbative point of view, the bridge can be
constructed to connect the Dirac Hamiltonian in realistic nuclei
with the symmetry limit of $S-V={\rm const}$, but not $S+V={\rm
const}$. This indicates that the realistic system can be treated as
a perturbation of the spin-symmetric Hamiltonian. This also confirms
in an explicit way that the nature of PSS is non-perturbative, as
the Dirac Hamiltonian with $S+V={\rm const}$ is regarded as the
symmetry limit.


However, it has been pointed out that the energy splitting of the
pseudospin doublets in realistic nuclei could be alternatively
considered as the breaking of their degeneracy appearing in the
spin-symmetric Hamiltonian with RHO potentials
\cite{Typel2008,Marcos2008}. In the following, we assess this
statement in a perturbative way.

In the case of the spin U(3) symmetry limit shown in
Ref.~\cite{Ginocchio2005PRL}, the Dirac Hamiltonian $H$ in
Eq.~(\ref{EQ:H}) is split as
\begin{equation}
    H = H_0^{\rm RHO} + W^{\rm RHO},
\end{equation}
with
\begin{equation}
H_0^{\rm RHO}=
    \lb\begin{array}{cc}
        \Sigma_{\rm HO}+M & -\frac{d}{dr}+\frac{\kappa}{r} \\
        \frac{d}{dr}+\frac{\kappa}{r} & -\Delta_0-M
    \end{array}\rb,
\end{equation}
and
\begin{equation}
    W^{\rm RHO} =
    \lb\begin{array}{cc}
        \Sigma-\Sigma_{\rm HO} & 0 \\
        0 & \Delta_0-\Delta
    \end{array}\rb,
\end{equation}
where $\Sigma_{\rm HO}(r) = c_0 + c_2 r^2$ has the form of a
harmonic oscillator. Here, $H_0^{\rm RHO}$ leads to the energy
degeneracy of the whole major shell, and $W^{\rm RHO}$ is identified
as the corresponding symmetry-breaking potential. In the present
investigation, we choose $-\Delta_0-M=-350$~MeV as in $H_0^{\rm SS}$
and $c_0+M=865$~MeV. As discussed before, the perturbative
properties are not sensitive to these two constants. Meanwhile, the
coefficient $c_2$ is chosen as 1.00~MeV/fm$^2$ to minimize the
perturbations to the $pf$ states.

In the upper panels of Fig.~\ref{Fig4}, the values of $\left|W^{\rm
RHO}_{mk}/(E_m-E_k)\right|$ for the pseudospin doublets
$k=1\tilde{d}$ are shown as functions of the energy differences
$E_m-E_k$. It is found that the general patterns shown in panels (a)
and (b) are the same as those in panels (a) and (b) of
Fig.~\ref{Fig2}, respectively, and the largest perturbation
correction is roughly 0.16 ($H_0^{\rm RHO}$ to $H$) or 0.10 ($H$ to
$H_0^{\rm RHO}$). This indicates that the criterion in
Eq.~(\ref{EQ:criterion}) is fulfilled, even though not as well as in
the SS case. In the lower panels of Fig.~\ref{Fig4}, the
single-particle energies of the states in the $pf$ major shell
obtained at the exact symmetry limit and by the self-consistent RMF
theory, as well as their counterparts obtained in the first-,
second-, and third-order perturbation calculations, are shown. As
shown in panel (c), not only the spin-orbit splitting, but also the
pseudospin-orbit splitting, is well reproduced by the third-order
perturbation calculations, and in the reversed way the energy
degeneracy of all the states in the $pf$ major shell can be well
restored as shown in panel (d). Thus, the link between the $pf$
states in realistic nuclei and their counterparts in the symmetry
limit with RHO potential can be explicitly established. Furthermore,
it is found that the single-particle wave functions of $H$
($H_0^{\rm RHO}$) can also be reproduced by the second-order
perturbation calculations starting from $H_0^{\rm RHO}$ ($H$).

Therefore, the bridge connecting the Dirac Hamiltonian in realistic
nuclei and that with RHO potentials can be clearly constructed using
perturbation theory. This indicates that the energy splitting of the
pseudospin doublets can be regarded as a result of small
perturbation around the Dirac Hamiltonian with RHO potentials, where
the degeneracy of the pseudospin doublets appears.

Of course, one must always keep in mind that the actual picture in
nuclear spectra is generally more complex than that of a simple
potential model. Still, there are well-identified cases where the
core polarization effects are small enough (correlation and
polarization diagrams may compensate each other and give relatively
large spectroscopic factors) to allow for the notion of
single-particle state to hold.

In summary, the symmetries of the Dirac Hamiltonian and their
breaking in realistic nuclei are investigated in the framework of
perturbation theory. The present framework can provide a clear and
quantitative way for investigating the perturbative nature of SS and
PSS. By examining the perturbation corrections to the
single-particle energies and wave functions, the link between the
single-particle states in realistic nuclei and their counterparts in
the symmetry limits has been established. It is found that the
symmetry limits of $S-V={\rm const}$ and RHO potentials can be
connected to the Dirac Hamiltonian in realistic nuclei by
perturbation theory, but not $S+V={\rm const}$. In other words, it
is suggested that the realistic system can be treated as a
perturbation of spin-symmetric Hamiltonians, and the energy
splitting of the pseudospin doublets can be regarded as a result of
small perturbation around the Hamiltonian with RHO potentials, where
the pseudospin doublets are quasidegenerate.

The present investigation is based on simple RMF theory with only
scalar and vector potentials. It would be interesting to study the
corresponding symmetry limits in systems with non-local potentials
or tensor interactions such as those encountered in relativistic
Hartree-Fock approaches \cite{Long2006,Long2007,Liang2010}. The
analysis done in this work is easy to be generalized to such
investigations.

This work is partly supported by State 973 Program 2007CB815000, the
NSFC under Grant No. 10975008, and China Postdoctoral Science
Foundation No. 20100480149. One of the authors (H.L.) is grateful to
the French Embassy in Beijing for the financial support for his stay
in France.


\end{CJK*}
\end{document}